\documentclass{mpe_report}
\usepackage{graphicx,epsfig,color}

\begin{document}
\title{Pulsar Spin, Magnetic Fields, and Glitches}
\author{M. Ruderman} 
\institute{Department of Physics and Columbia Astrophysics Laboratory,  Columbia University, New York, NY}
\maketitle

\begin{abstract}
In the core of a canonical spinning magnetized neutron star(NS) a nearly uniform superfluid neutron vortex-array interacts strongly with a twisted array of magnetic flux-tubes threading the core's superconducting protons.
One consequence is that changes in NS-spin alter both arrays and also the magnetic field distribution on the surface of the surrounding crust.  Among predicted consequences for very young spinning-down NSs are ``spin-down indices" increasing from 2 to 3, and a family of (Crab-like) spin-period ``glitches" with permanent fractional jumps in spin-down torque 10$^5$ times greater than those in NS-spin. For older NSs, average spin-down indices increase to around 5, and an additional  (Vela-like) family of giant glitches develops.  NS spin-up to millesecond pulsars results in a high abundance of orthogonal and aligned rotators, and anomolously small polar cap areas.  Observations do not conflict with these expectations.   An epoch of NS magnetic field evolution between the onset of proton superconductivity ($\sim$ yr) and neutron superfluidity 
($\sim 10^3$ yrs ?) may be important for large surface magnetic field changes and needs further study.  Observations generally considered evidence for NS precession seem to need reconsideration.
\end{abstract}

\section{Introduction}

In a cool core below the crust of a spinning neutron star (NS) superconducting protons coexist with much more abundant superfluid neutrons (SF-n) to form a giant atomic nucleus which contains within it a neutralizing sea of relativistic degenate electrons.  Superfluid neutrons in a star with a spin-period $P (sec) \equiv 2\pi/\Omega$ rotate by forming a nearly uniform array of corotating quantized vortex lines parallel to the spin axis, with an area density 
$n_v  \sim 10^4~ P^{-1}$ cm$^{-2}$ (Fig. 1).

\begin{figure}
\includegraphics[width=4.4cm]{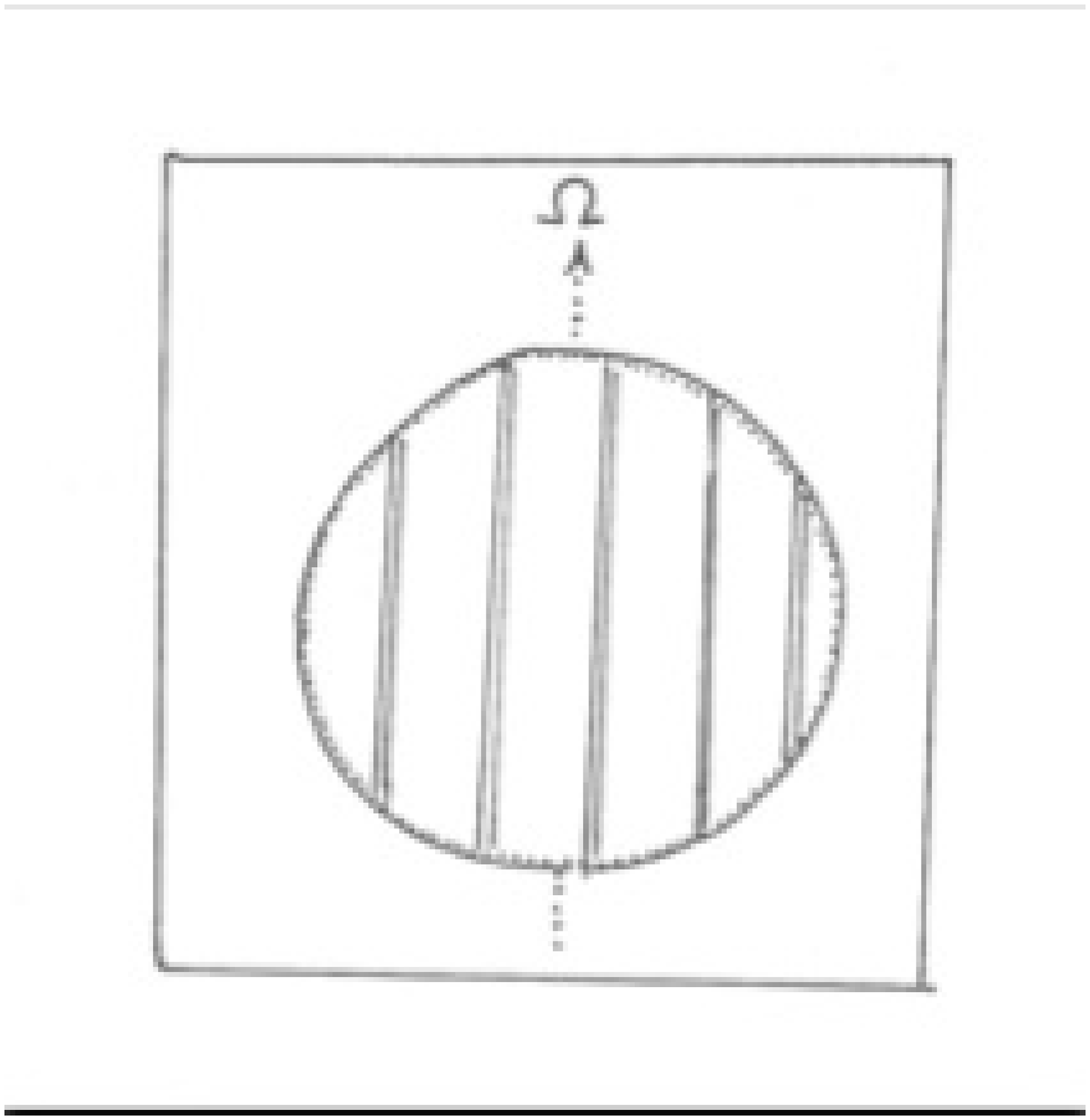}
\includegraphics[width=3.9cm]{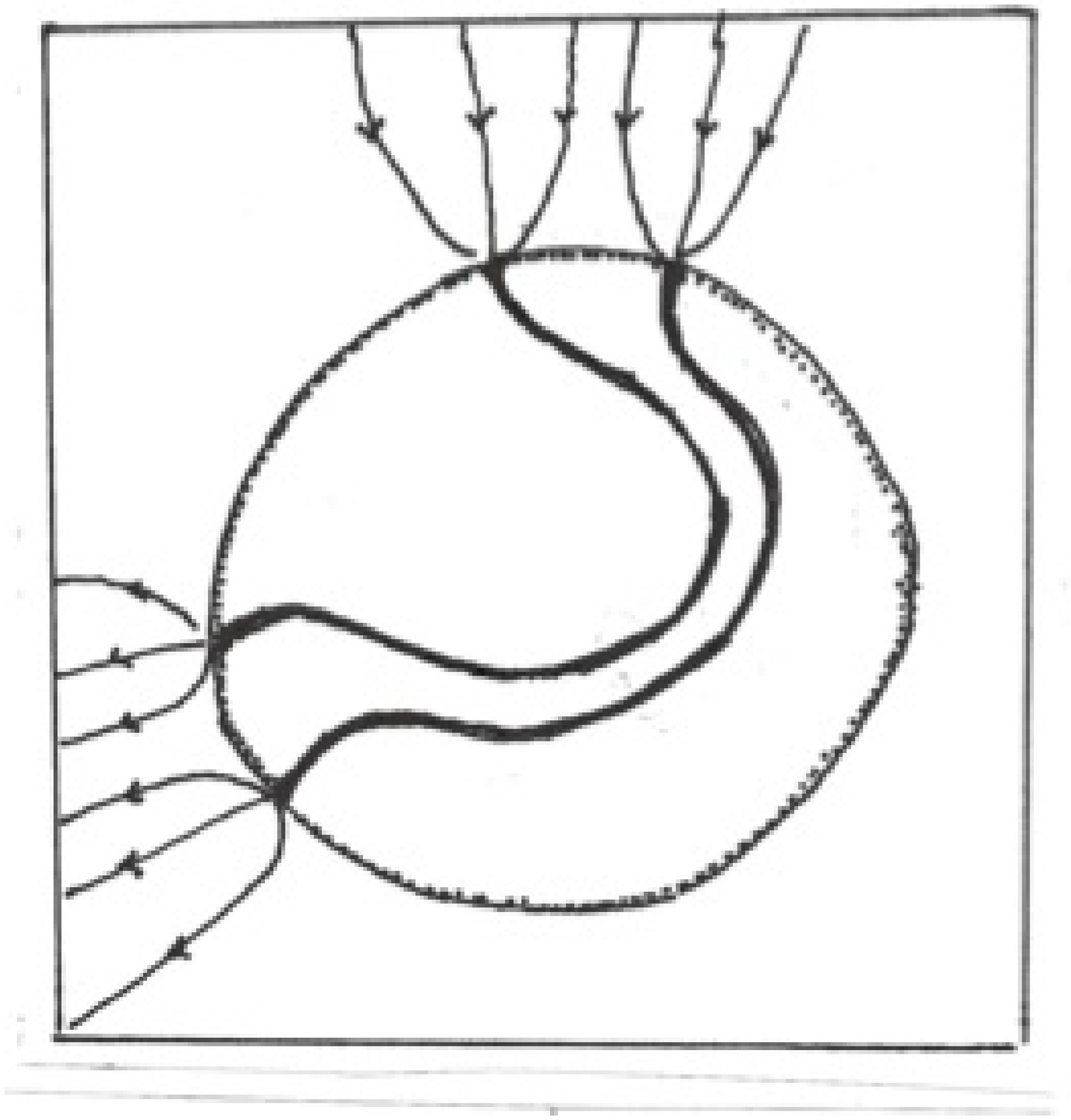}
\caption{(L) Five of the 2 R$^2 \Omega M_n/t~~ (\sim 3 \times 10^{16}/P$ (sec) n-sF vortex lines in the superfluid core of a cooled NS.}
\end{figure}

\noindent
The array must contract (expand) when the NS spins up (down).  In stellar core neutron spin-up or spin-down, a vortex a distance $r_\perp$ from the spin-axis generally moves outward with a velocity $r_\perp (\dot{P}/2P)$ until $r_\perp$ reaches the core's neutron superfluid radius (R).  Any stellar magnetic field passing below the stellar crust must, in order to penetrate through the core's superconducting protons (SC-p), become a very dense array of quantized flux-tubes 
($n_\phi \sim$ 5 $\times$ 10$^{18}$ B$_{12}$ cm$^{-2}$ with $B$ the local average magnetic field).  Each tube carries a flux of 2 $\times$ 10$^{-7}$ G cm$^2$
and a magnetic field $B_c \sim 10^{15}$ G.\footnote{This assumes Type II proton-superconductivity in the NS core below the crust, the common result of many calculations.  If it were Type I, with many thin regions of $B >$ several $B_c$, and $B \sim$ 0 in between ({Link 2003}), the impact on surface $B$ of changing NS spin proposed below would not change significantly.   If,  however,  the locally averaged $B$ inside the NS core exceeds a critical field somewhat greater than $B_c$, the core's protons would not become superconducting.  This may well be the case for most (or all) ``Magnetars".}   Tension along flux tube bundles $\sim BB_c/8 \pi$ but there is negligible intervention between flux tubes as long as $B << B_c$.  The initial magnetic field within the core of a neutron
star is expected to have both toroidal and very non-uniform poloidal components.   The web of flux-tubes formed after the
transition to superconductivity is then much more complicated and irregular than the neutron vortex-array as well as of order $10^{14}$ times more dense (Fig. 2).

\begin{figure}
\caption{(R) Two of the 6 $\cdot 10^{31} \rm{B}_{12}$ magnetic flux-tubes in the superconductivity  core
of a NS.   Each flux-tube has a flux $2 \times 10^{-7} \rm{G~cm}^2$ and a magnetic field B$_c \sim 10^{15}$ G. }                                                                             
\end{figure}

Because of the velocity dependence of the short range nuclear force between neutrons and protons, there is a strong interaction between the neutron-superfluid's vortex-lines and the proton-superconductor's flux-tubes if they come closer to each other than about $10^{-11}$ cm.  Consequently, when  $\dot{P} \neq 0$,  flux-tubes will be pushed (or pulled) by the moving neutron vortices 
({Sauls 1989;  Srinivasan et al. 1990;  Ruderman 1991;  Ding et al. 1993; Ruderman et al. 1998; Jahan-Miri 2000;  Konenkov \& Geppert 2001;  Ruderman 2005})
(Fig. 3).
A realistic flux-tube array  will be forced to move along with a changing SF-n vortex array which threads it as long as the force at a vortex-line flux-tube junction does not grow so large that vortex-lines cut through flux-tubes.  In spinning-down pulsars cold enough to have SF-n cores (T$_{core}$$^<_\sim$ 3 $\times$10$^8$ K) outward moving n-vortex velocities are generally

\begin{figure}
\centerline{\includegraphics[width=4.4cm]{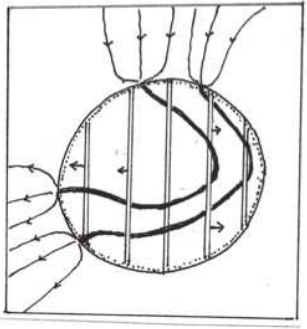}}
\caption{Interacting flux-tubes and vortex-lines during initial spin-down.}
\end{figure}

\noindent
less than the cm/day in the Crab pulsar.  During NS spin-up to millisecond pulsars the inward moving n-vortex velocities are usually $<$ cm/century.  The drag on flux-tubes caused by such slow movement is far too small to cause cut-through of flux-tubes by moving n-vortices.  ({Jones 2006}) has recently found that electron scattering on flux-tube cores allows easier passage of flux-tubes through the SC-p than had been estimated previously. (In addition, an expected motion-induced flux-tube bunching instability would more easily allow easy co-motion of flux-tubes with the local electron plus SC-p fluid in which they are embedded ({{Ruderman 2004a}). )
If not for the anchoring of flub-tubes at the base of the metallic crust (idealized in Fig. 4a) flux-tube positions at the core-crust interface could  closely follow changes in the core's SFn-vortex array.  

The magnetic field at the crust-surface would always be almost equal to the average field from the flux-rubes at the core-crust interface below on time-scales exceeding the ``impurity"-dominated Eddy diffusion time through the crust, estimated to be 
several million yrs.    On much shorter time-scales the crust acts like a thin, incompressible, solid, breakable, perfect conductor.  Because of crustal stratification and the huge gravitational field  it yields mainly to huge shear-stresses (from $B B_c/8 \pi \sim 3 B_{12}$ tons/cm$^2$)  larger than its yield strength.   This happens on scales L ($>$ crust thickness  $\Delta$; Fig. 4b)  which satisfy

\begin{equation}
\frac{BB_c}{8 \pi}~L^2~^>_\sim~(\mu~\theta_m \Delta) L~~.
\end{equation}

\noindent
Then

\begin{equation}
L ~^>_\sim~\frac{10^6 \rm{cm}}{B_{12}}~\sim~ \frac{R}{B_{12}}
\end{equation}

\begin{figure}
\includegraphics[width=8.8cm]{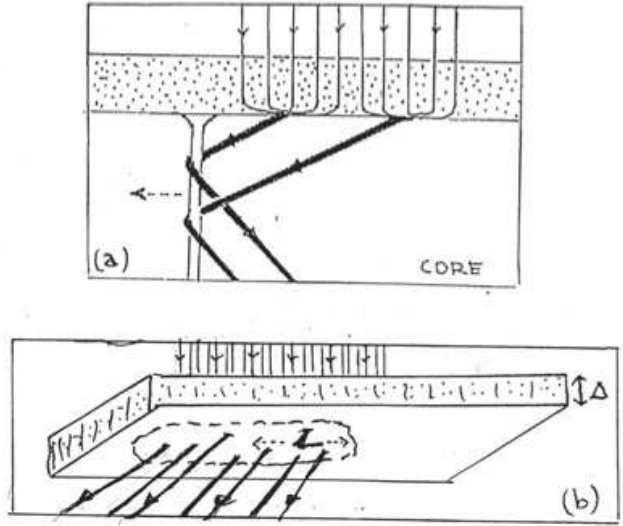}
\caption{(a) A moving quantized vortex-line in a NS core's superfluid neutrons puling a pair of the core's proton
superfluid quantized flux-tubes anchored in the star's solid, conducting crust (shown dotted).                                                   
(b) The scale L of Eq (1) for the largest shearing stress on the crust from flux-tube pull.}                                    
\end{figure}

\noindent
for typical estimates of  crust shear modulus $\mu \sim$ 10$^{30}$ dyne cm$^{-2}$, $B_c \sim 10^{15} G$,
$\Delta \sim 10^5$ cm, and maximum sustainable stain  $\theta_m \sim 3 \times  10^{-4}$.
This forms the basis for a very simple model for describing predicted changes in pulsar magnetic fields during NS spin-up or spin-down which agrees well with  different families of pulsar observations.
On small scales ($<$ L) the magnetic field through the crust of a NS can be frozen for the first several 10$^6$ yrs after the crust solidifies (several hours after the formation of the NS).   It changes  during this long epoch  only when huge
$BB_c/8 \pi$ shearing stresses overstrain the crust on large scales ($>$ L), inducing  ``platelets" with different B to interchange positions by slow ``plastic flow" or more sudden discontinuous crust-breaking.   After the formation of n-SF vortex arrays
($t ^>_\sim 10^3$ yrs) large scale magnetic fields and dipole moments follow underlying core n-vortex movement but local polar cap field distributions
in small areas (a typical polar cap radius $\sim 10^4$ cm) do not change substantially because of such movement.  However, after several $10^6$ yrs both surface dipole moments and surface polar cap field strength to follow  closely the  movement of SF-n vortices near the top of the NS core.

\section{Magnetic field changes in spinning down neutron stars}

Effects from the coupling between a spin-down expansion of a NS's SF-n vortex-array and its SC-p flux-tubes should appear in several observable phases after a neutron star spin-period $P_o$ is reached when the NS has cooled  enough that the vortex-line array and the flux-tube one have both formed (10$^3$ yrs).

The distance from the spin-axis of nSF vortex lines ($r_\perp$) increases with decreasing NS spin  ($\Omega$) to conserve 
$\Omega r_\perp ^2$.   This has the following consequences.
\vspace*{8pt}

\noindent
a)~~ When $P > P_o, ~\mu_\perp$, the component of magnetic dipole moment perpendicular to NS spin, initially grows as $P^{1/2}$ for any configuration of surface B (cf. Fig. 5):

\begin{equation}
\frac{\mu_\perp (P)}{\mu_\perp (P_o)}  \sim \left( \frac{P}{P_o} \right)^{1/2}
\end{equation}

\vspace*{8pt}

\noindent
b)~~ When $P \sim$ several P$_o$, a good fraction of a NS's core flux -tubes will have been pushed outwards from the spin-axis into the equatorial region $r_\perp \sim R$.   These cannot, of course, continue to move outward (Fig. 5) so that Eqn (3) no longer holds.  Rather, the mixture of expanding and crust-constrained flux-tubes gives:

\begin{equation}
\frac{\mu_\perp (P)}{\mu_\perp (P_o)} \sim \left( \frac{P}{P_o} \right)^{\hat{n}}~~(0 < \hat{n} < 1/2)
\end{equation}

\noindent
with the exact value of $\hat{n}$  dependent on details of the core's $B$-field configuration.
\vspace*{8pt}

\begin{figure}
\includegraphics[width=8.8cm]{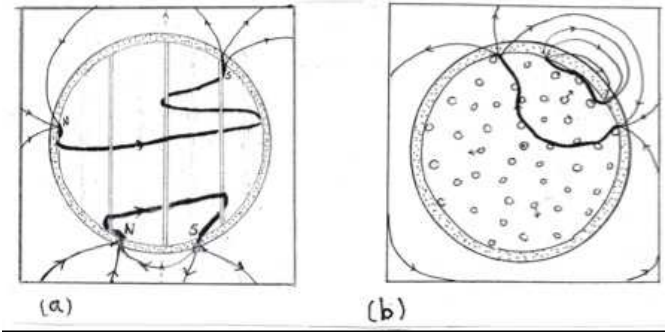}
\caption{Two flux-tubes whose North and South magnetic poles moved away from a NS's spin-axis during spin-down:  (a) side view with one of the North poles moved to its maximum  $r_\perp \sim$ R;  (b) top view of the core's equatorial plane with one of the flux tubes being pushed out of the NS core.}                                                                            
\end{figure}

\noindent
c)~~ The crust can delay, but not indefinitely prevent, expulsion of this magnetic field from the NS.  When $P \sim$ several P$_o$, intertwined vortex plus flux which have been pushed into the core-crust interface will stress the crust enough to exceed its shear-strength (Sect. 5. and Figs. 6, 19 \& 20).   Then crust movements begin that lead to $B$-field reconnections.
Flux that is threaded by SF-n vortex lines that have not yet reached $r_\perp \sim R$,
and thus have not yet disappeared in this way are the remaining source for the NS's dipole moment.  The sum of all this remaining flux $\propto$ the total number of remaining vortex-lines ($\propto \Omega$).    Then, Eqn. 4 holds with $\hat{n} = -1$.
\vspace*{8pt}

\noindent
d)~~ When the $B$ remaining in the core ($\propto \Omega)$ drops to and below about  10$^{12}$ G, shear-stress on the crust would no longer be expected to exceed the crust's yield-strength.  The NS's surface $B$ may then lag that at the base of its crust by as much as $10^7$ yrs., slightly greater than the crust's Eddy current dissipation time.

\section{Magnetic dipole field changes in spinning-up neutron stars}

NS spin-up, when sustained long enough so that one of the above criteria for limiting shear-stress from crust-anchored magnetic flux before cut-through is met,  leads to a ``squeezing"  of surface {\bf B} toward the NS spin-axis.    After a large decrease in spin-period from an initial $P_o$ to $P \ll P_o$ all flux would enter and leave the core's surface from the small area within a radius $R(P/P_o)^{1/2}$ of the NS's spin-axis.   This  surface {\bf B}-field change is represented in Figs. 6-7 for the special case when the magnetic flux which exits the NS surface from its upper (lower) spin-hemisphere returns   to the stellar surface in its lower (upper) one.   Potentially observable features of such a ``spin-squeezed" surface {\bf B} configuration  include the following.

\begin{enumerate}
\item[a)] The  dipole moment  becomes nearly aligned along the NS spin-axis.
\vspace*{8pt}

\item[b)] The canonical polar cap radius, $r_p \equiv R (\Omega R/c)^{1/2}$, shrinks to 
$r^\prime_p \equiv \Delta (\Omega R/c)^{1/2}$.
The $B$-field just above the polar cap  has almost no curvature.
\end{enumerate}
\vspace*{8pt}

\begin{figure}
\includegraphics[width=4.2cm]{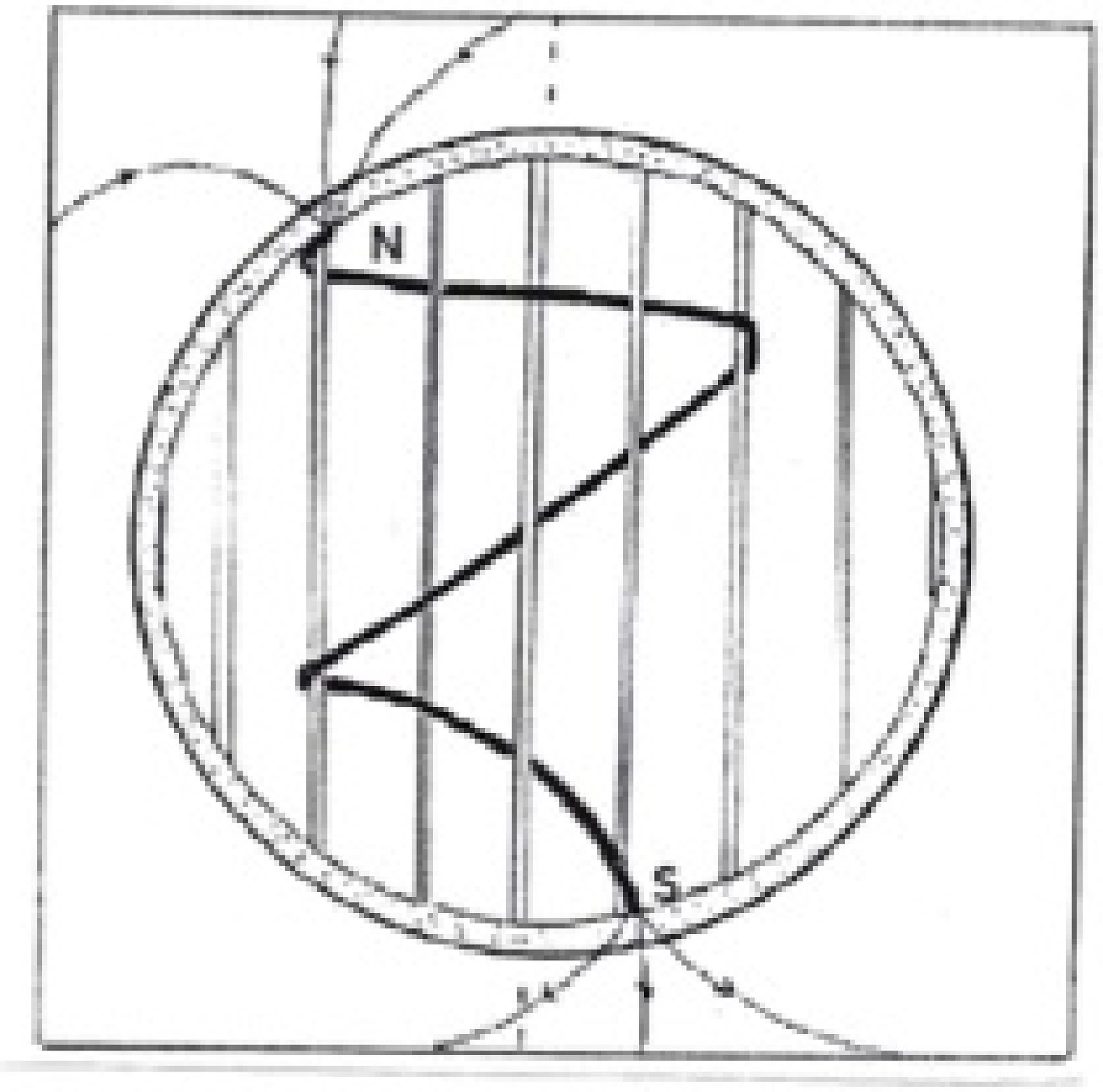}
\includegraphics[width=4.2cm]{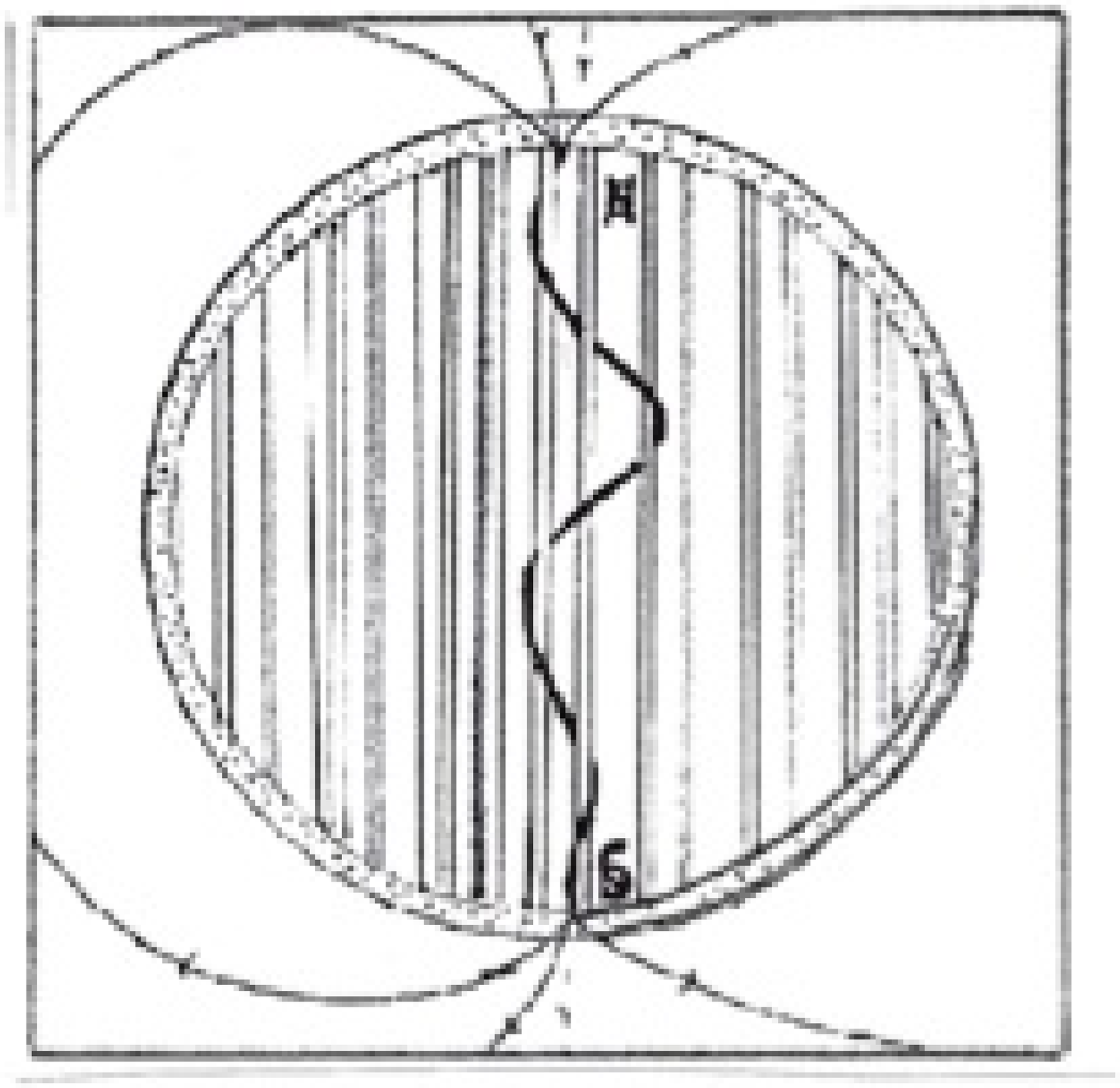}
\caption{(L) A single flux-tube (one of $10^{31}$) and some of the NS's arrayed vortices
(8 of $10^{17}$) with surface poles in opposite spin-hemispheres.}                                     
\end{figure}

\begin{figure}
\caption{(R) The flux-tube and vortex array of Fig. 6 after a large stellar spin-up.}                                                 
\end{figure}

\noindent
If the pre-spin-up surface $B$ has a sunspot-like configuration (i.e. flux returning to the NS surface in the same hemisphere as that from which it left), the spin-up-squeezed field change is represented in Figs. 8 \& 9.   In  this case, potentially observable features when $P \equiv P_o$ include the following:

\begin{enumerate}
\item[c)] A pulsar dipole moment nearly orthogonal to the NS spin-axis, and 
\vspace*{8pt}

\item [d)] positioned at the cruse-core interface.
\vspace*{8pt}

\item[e)] A dipole moment $\mu$ reduced  from its pre-spin-up size (as in Eq (1)) with $P \equiv P_o$.
\end{enumerate}

\begin{figure}
\includegraphics[width=4.2cm]{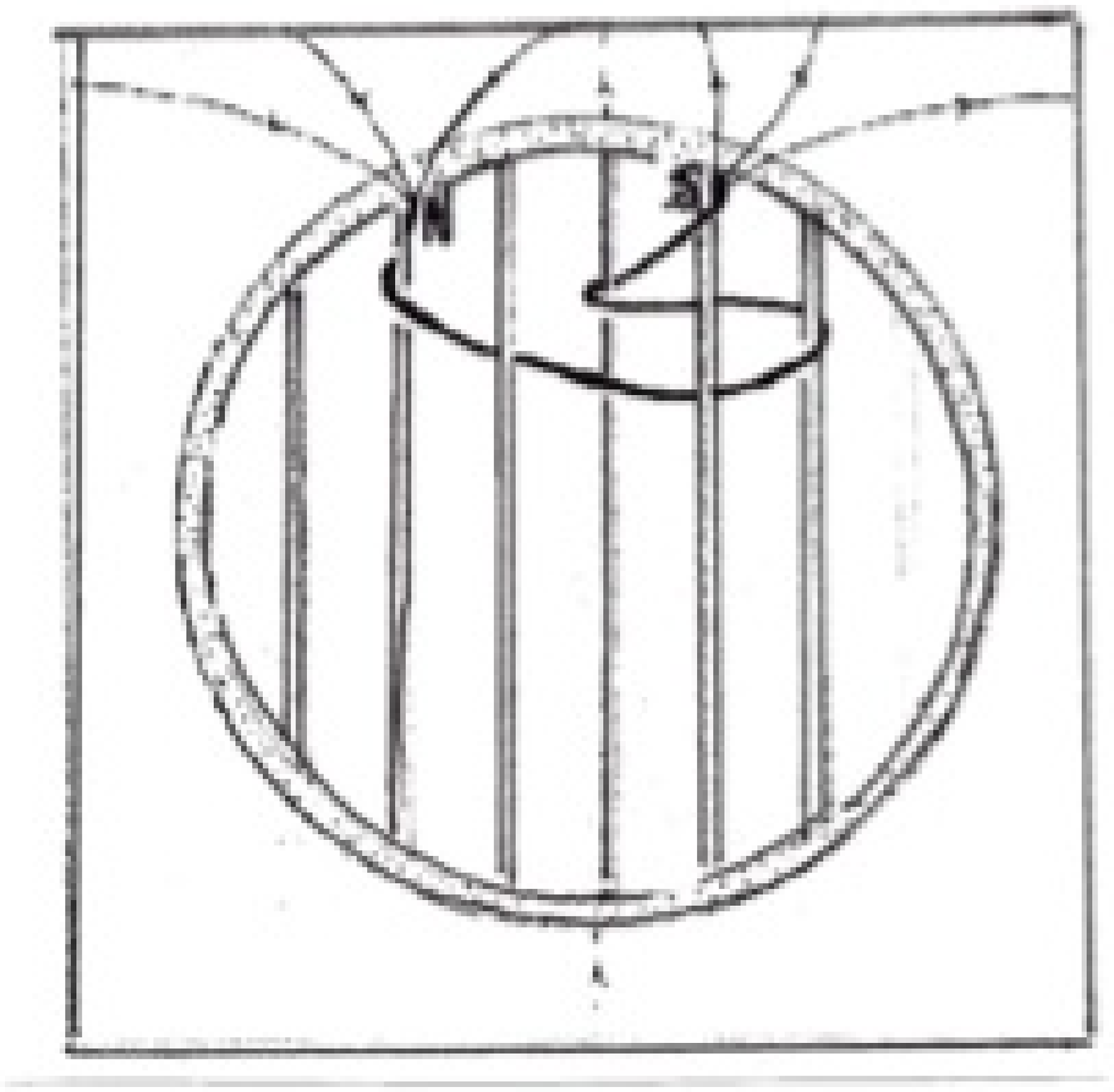}
\includegraphics[width=4.2cm]{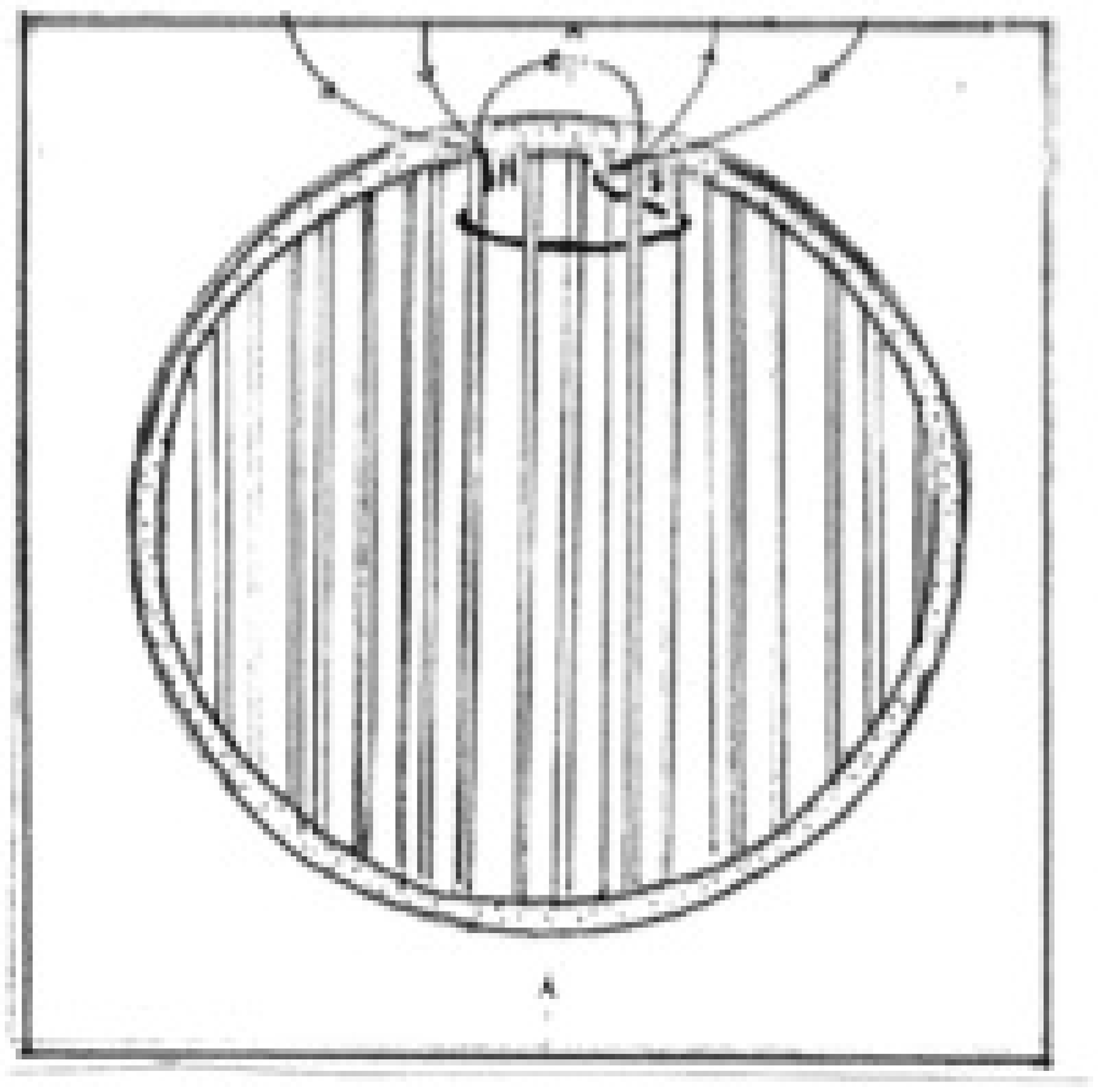}
\caption{(L)  A single flux-tube, part of a sunspot-like $B$-field geometry in which flux from a spin-hemisphere of the surface returns to the surface in that same hemisphere.}                                     
\end{figure}

\begin{figure}
\caption{(R) The flux-tube and vortex array of Fig. 8 after a large stellar spin-up.}                                      
\end{figure}

\noindent
A more general (probably more realistic) pre-spin-up configuration has flux emitted from one spin-hemisphere returning to the stellar surface in both, as in Fig. 10.  Spin-up squeezing then typically gives the surface field configuration in Fig. 11, a spin-squeezed, nearly orthogonal dipole on the NS  spin-axis with properties (d), (e), and (f), together with an aligned dipole on the spin-axis whose external field is well-represented by North and South poles a distance  2R  apart.  Further spin-up could lead to the Figs. 12  and  7 configuration; that of Fig. 13 and 9 would be realized only if $S_2$ of Figs. 10 and 11 is negligible.

\begin{figure}
\includegraphics[width=4.2cm]{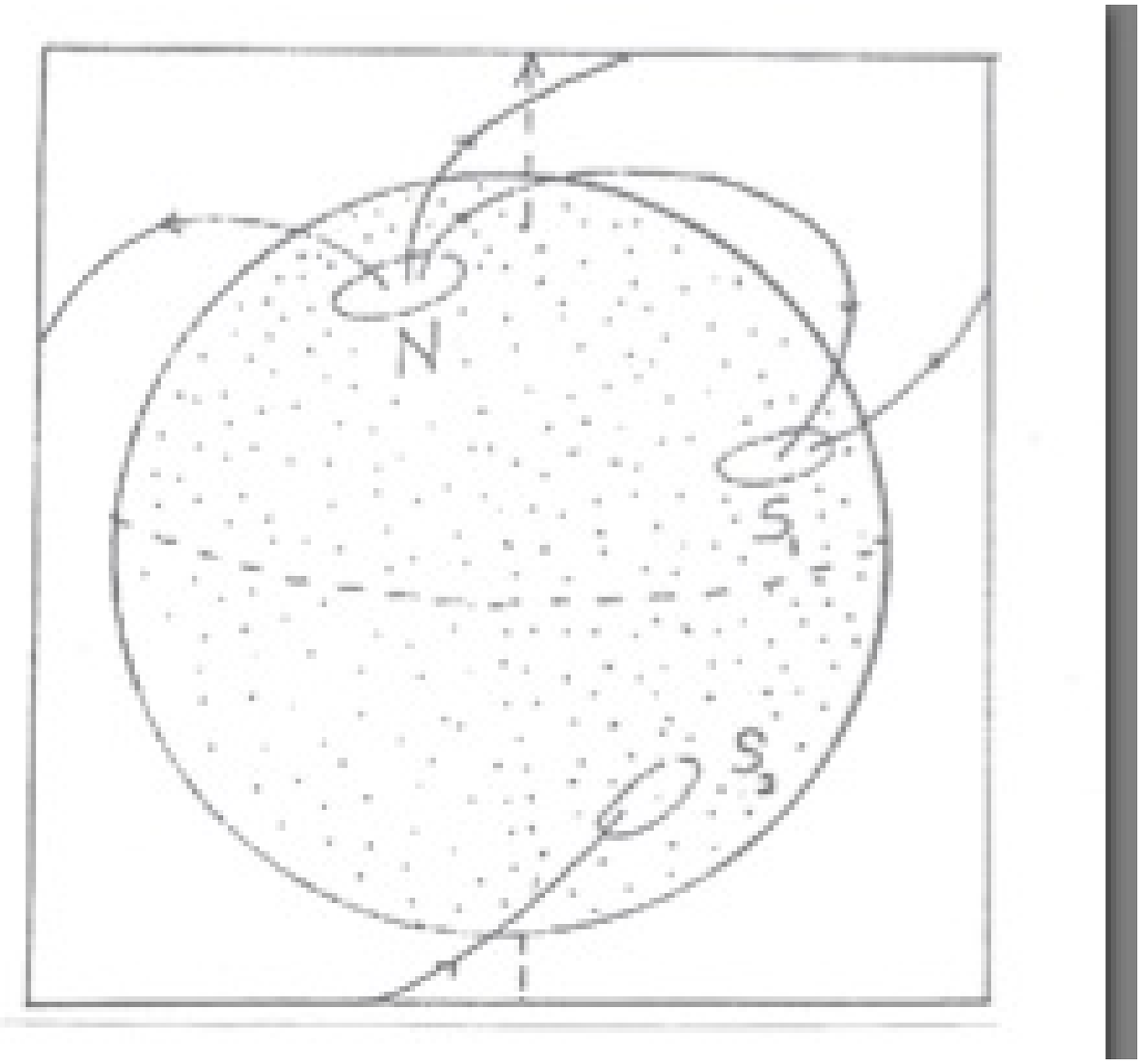}
\includegraphics[width=4.2cm]{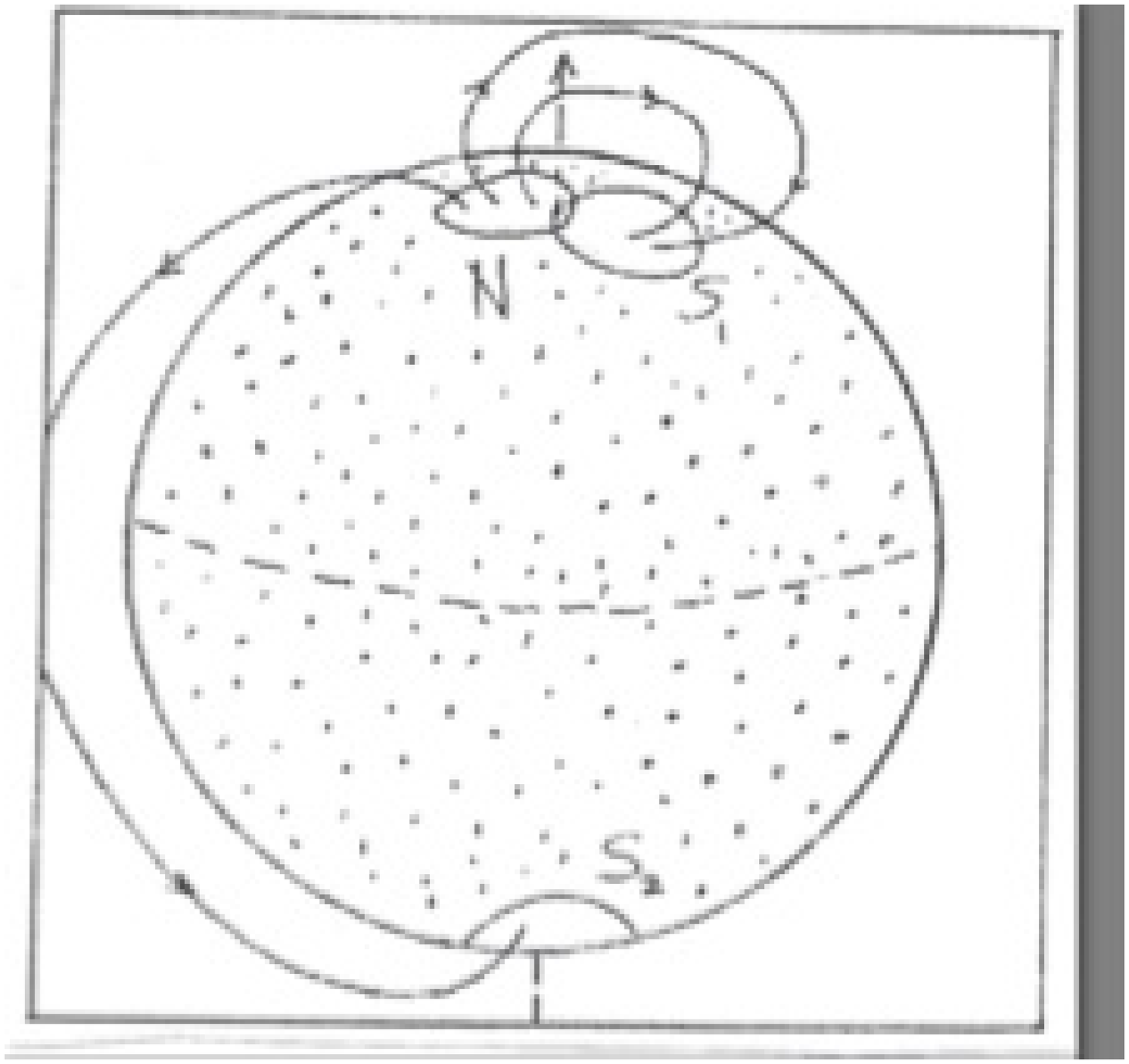}
\caption{(L) A surface field which has flux of both Fig. 6 and Fig. 8 configurations.}                                                   
\end{figure}

\begin{figure}
\caption{(R) The field from Fig. 10 after a large stellar spin-up.}                                                            
\end{figure}

\begin{figure}
\includegraphics[width=4.2cm]{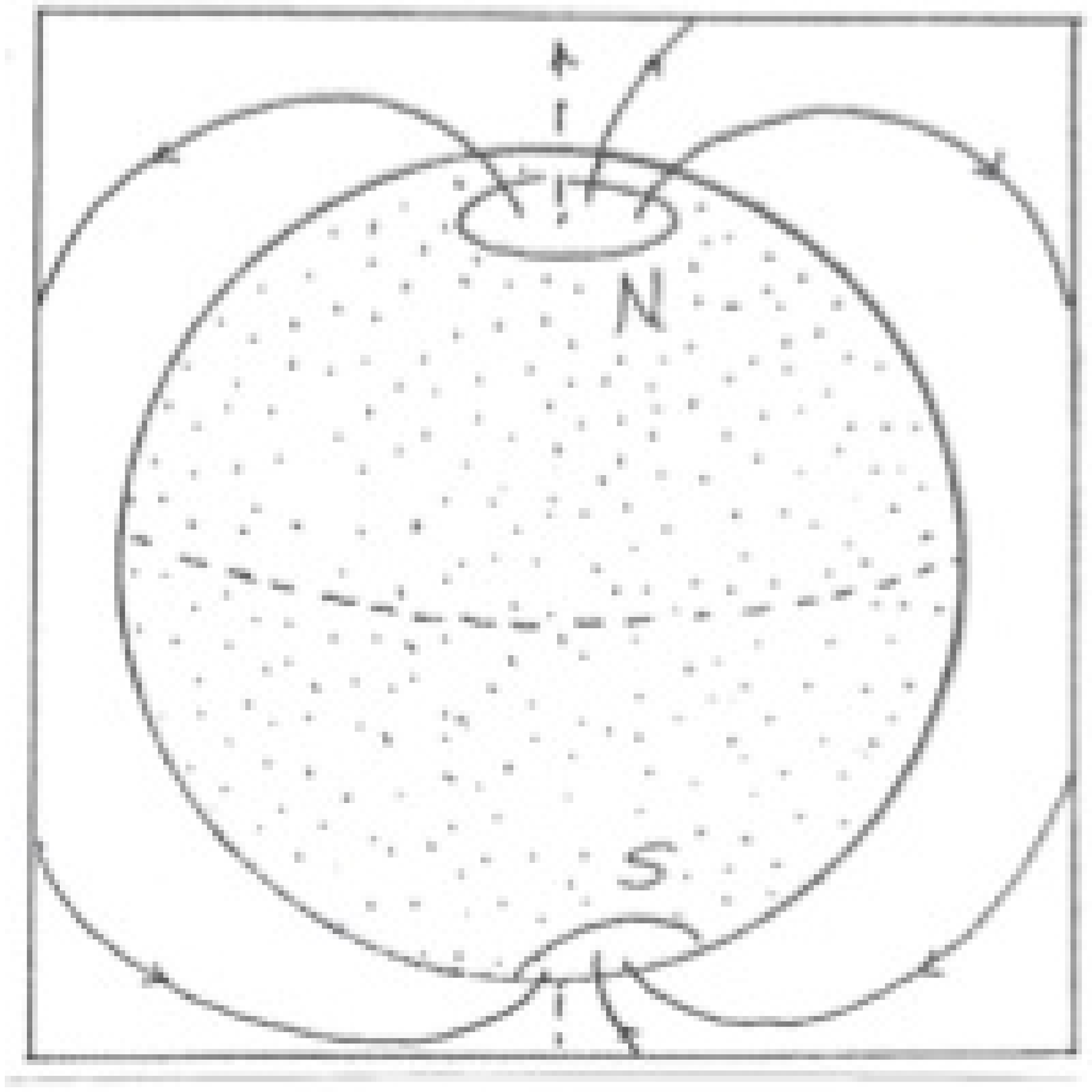}
\includegraphics[width=4.2cm]{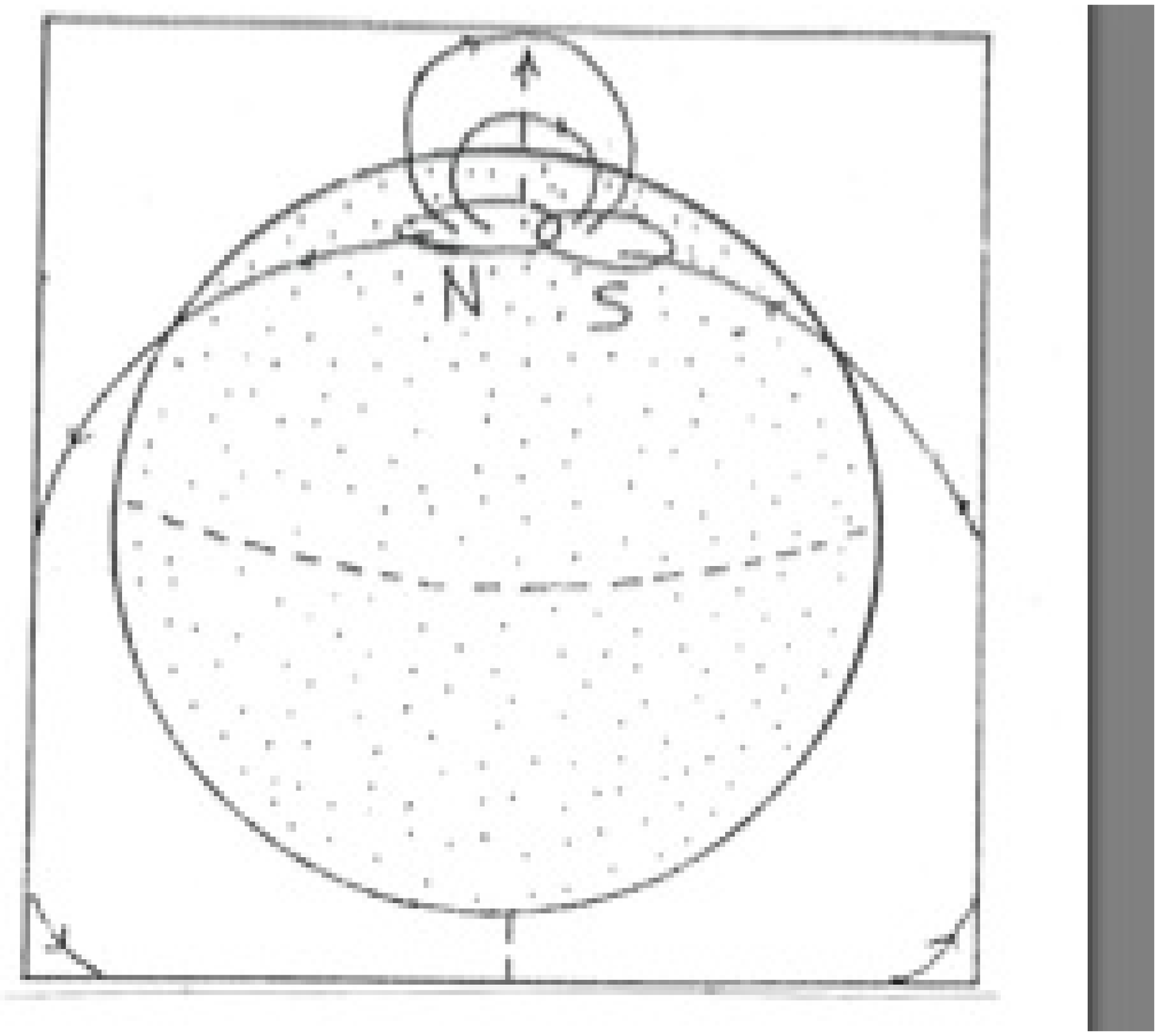}
\caption{(L) The field from Fig. 11 after further spin-up.}                                      
\end{figure}

\begin{figure}
\caption{(R) The field from Fig. 10 after large spin-up when the $S_2$ contribution to Figs. 10 and 11 is negligible.}                                                                                                                   
\end{figure}

\section{Comparisons of pulsar dipole field observations with model expectations}

Fig. 14 shows observationally inferred surface dipole fields $(B)$ as a function of  $P$ for about $10^3$ radiopulsars ($B$ is calculated from measured $P$ and $\hat{P}$, I$\dot{\Omega} = -\mu^2_\perp \Omega^3 c^{-3}; B = \mu_\perp R^{-3}$ and $I = 10^{45}$ g cm$^2$.).   Segments of $B$(P) based upon the model of Sects 2 and 3, are shown for a typical pulsar by the doubled or single solid lines.

\begin{figure}
\includegraphics[width=8.8cm]{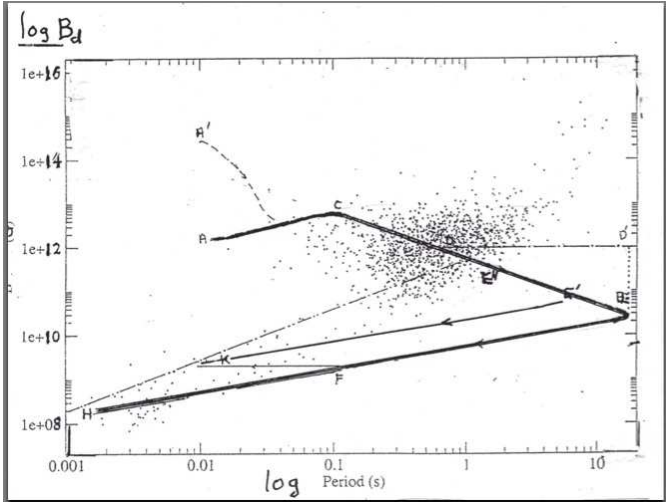}
\caption{Dipole-$B$ observed on pulsar surfaces (inferred from measured $P$ and $\dot{P}$) as a function of pulsar period ($P$) (12).
The solid line segments are the evolutionary segments for $B$ of Sect. 4, based upon the model of Sects. 2 and 3.  The dash-dot diagonal is the steady state spin-up line from an accretion disk fed at 0.1 the Eddington limit.  The horizontal ($D \rightarrow D^{\prime}$) is that for a NS surface above core surface ($D \rightarrow E$).}                                                                                                                                 
\end{figure}

\begin{enumerate}

\item Point A is the ($B,P$) where, typically, flux-tubes and vortex lines begin coexistence.
\vspace*{8pt}

\item ($A \rightarrow C)$ is the expanding array prediction of
Sect. 2(b): $B \propto P^{\hat{n}}$ with the model prediction $0 <
\hat{n} < 0.5$.  The index $\hat{n}$ is known only in the several
cases where $\ddot{P}$ is also measured: $\hat{n}$ = + 0.3, 0.1, 0.6,
0.1, 0.05 ({Lyne et al. 1998; Kaspi et al. 1994, Zhang \& Kojima
2005; Camilo et al. 2000}).  (The canonical ``spin-down index" $ n =
3-2 \hat{n}$)

\item $(C \rightarrow D)$ is the flux-expulsion and continual
reconnection segment of Sect. 2(e).  The model predicts $< \hat{n}>$ =
-1 for $\hat{n}$ averaged over the $(C \rightarrow D)$ history of any
one pulsar.  Reliable $\ddot{P}$ are not generally measurable in this
($B, P$) region.  However comparison of the spin-down times $P/2
\dot{P}$ with actual pulsar ages inferred from probable distance
traveled since birth ({Cordes \& Chernoff 1998}), gives $< \hat{n}> =
-0.8 \pm 0.4$, not inconsistent with the model prediction.
\vspace*{8pt}

\item $(D \rightarrow E)$ is the core-surface/crust-base $B$ evolution
for $\sim 10^{10}$ yrs.  The horizontal $(D \rightarrow D^{\prime})$
is the NS crust's surface field, remaining near $10^{12}$ G for $\sim
10^7$ yrs. as discussed in Sect. 2 (d).  This segment should be
characteristic of typical ``X-ray pulsars" (NSs in binaries spun up or
down by active companions through a wide range of $P$ (e.g. Hercules
X-1 with $P \sim$ 1s to Vela X-1 with $P \sim 10^3s$) until crustal
Eddy current decay allows a $(D^{\prime}\rightarrow E)$ decay from
some $D^{\prime}$ region.  A small minority of NSs, after $(D
\rightarrow E)$ segments, will be resurrected by accretion from a
previously passive White Dwarf companion which now overflows its Roche
lobe (LMXBs).  These NSs have entered into the spin-up phase of
Sect. 3 until they reach a steady state on the canonical ``spin-up
line" represented by the dot-dashed diagonal of Fig. 14 (for $M =
10^{-1} M_{Eddington})$.
\vspace*{8pt}

\item $(E \rightarrow F \rightarrow H)$ is the spin-up segment when
the NS surface $B$ has the sunspot geometry of Figs. 13 and 9, which
allows spin-up to minimal $P$ before spin-up equilibrium is
reached. Observations of maximally spun-up millisecond pulsars (MSPs)
support the Sec. 3 model for such MSP formation.  Figs. 15 and 16 show
Sect. 3 (d)'s high fraction of MSPs with two subpulses about
180$^\circ$ apart, characteristic of orthogonal rotators (Figs. 15 and
16) ({Cordes \& Chernoff 1998; Ruderman 2004b; Jayawardhana \&
Grindlay 1995; Chen \& Ruderman 1993; Becker \& Aschenbach 2002}).
Sect. 3 (e)'s $B$-field geometry is consistent with the linear
polarization and its frequency dependence in such subpulses ({Chen \&
Ruderman 1993}).

\begin{figure}
\includegraphics[width=8.8cm]{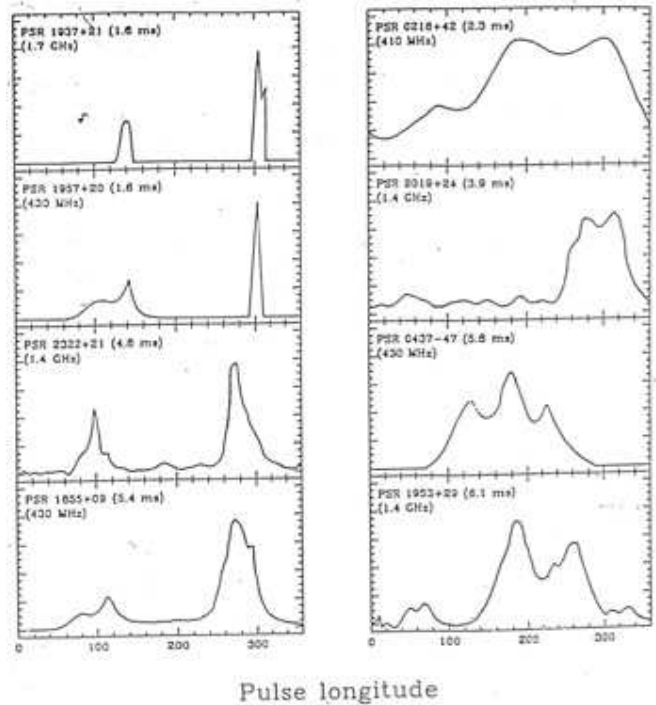}
\caption{Radiopulse profiles of eight of the fastest eleven disk
population millisecond pulsars in 1997 (4).  Among non-MSPs about one
pulsar in $10^2$ have two sub pulses near half a period apart.}
\end{figure}

\begin{figure}
\includegraphics[width=8.8cm]{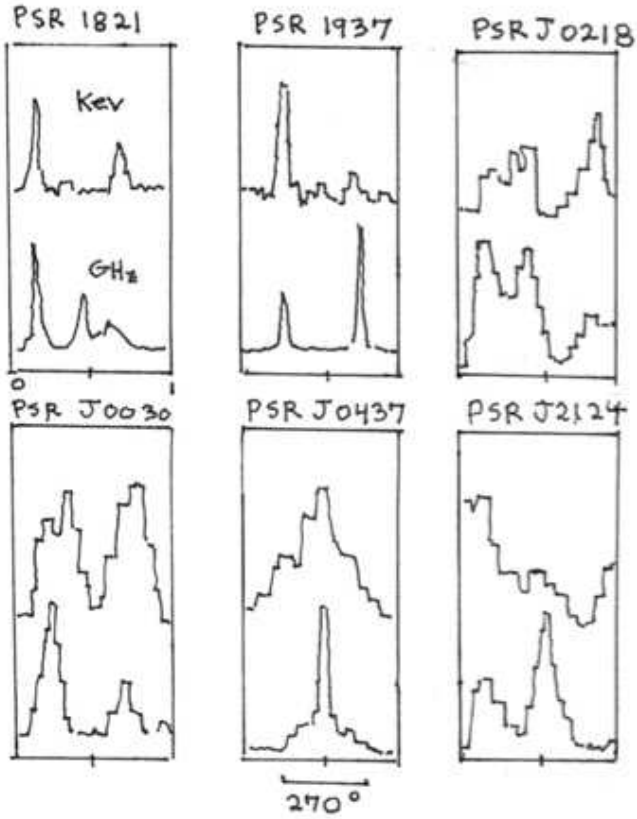}
\caption{The (GHz) radio pulse (lower segments) and keV x-ray (upper
segments) of the first six MSPs observed in x-rays (1).  PSR 0437 has
a pulse profile and width (270$^\circ$) which infer an almost aligned
radio pulsar observed from a direction not far from the NS spin axis.}
\end{figure}

\item $(E \rightarrow F \rightarrow K)$ is the track of surface $B$
(here the total dipole field) predicted after large spin-up from $(E)$
with Fig. 10 geometry to $(F)$ with Fig. 11 geometry.  Further spin-up
diminishes only the orthogonal component of $\mu$ until an almost
aligned rotator (Figs. 7 and 12) results when $(K)$ is reached.  X-ray
emission from the almost aligned MSP PSR 0437 ($P = 6$~ms) (cf
Figs. 15 and 16) supports the predicted tiny polar cap area about
$(\Delta/R)^2 \sim 10^{-2}$ that from a central dipole moment
configuration for the same $P$ (\S 3b and refs Chen \& Ruderman
1993; Ruderman 2004b; Becker \& Ashenbach 2002; cf Eq. 5 below).
\vspace*{8pt}

Expected consequences for pulsar diplole-{\bf B} changing according to
the Sects. 2-3 model and Fig. 14 are supported by many kinds of
observations.  However, for almost all there is usually another
popular explanation (e.g. $B$ getting from $(D)$ to $(H)$ just by
burial of {\bf B} by accreted matter from a companion (van den Heuvel
\& Bitzaraki 1995; Burderi \& D'Amico 1997; Zhang \& Kojima 2005).

\end{enumerate}

\section {Polar cap areas}

A NS's polar cap area $(A_{pc})$ depends upon both its dipole field $B_d$ and its polar cap field $(B_{pc})$:

\begin{equation}
A_{pc} \sim \frac{\Omega R^3}{c} \frac{B_d}{B_{pc}}
\end{equation}

\noindent
Since $B_{pc}$ is expected to be approximately constant for the first
few $ \times 10^6$ yrs after crust formation $A_{pc}$ $\infty B_d/P$
in that interval.  It is not clear, however, what the appropriate
initial value of $B_{pc}$ is.  In many cases it may well be very much
larger than $B_d$ after $10^3$ yrs when the controlling $n-SF$ vortex
arrays form and the model considered above can become applicable (cf
Sect. 5).

Very slow (t $>> 10^6$ yrs) spin-up of a NS to a millisecond pulsar  squeezes all the core's flux-tubes, together with the NS polar caps, toward the spin axis.  $A_{pc}$ then decreases both because $B_d$ becomes smaller ($B_d \propto \Omega^{1/2}$ on E  $\rightarrow$ F $\rightarrow$ H in Fig. 14) and $B_{pc}$ increases $B_{pc} \propto \Omega$).   Spin-up from a spin-period $\sim$ 10 s to near the minimum $P \sim 10^{-3} s$ on the 
``spin-up" line for near maximum NS spin-up by a companion-fed accretion disk gives a polar cap radius at the core-crust intercase which is $<<$ the crust thickness $\Delta$.     Then for an aligned rotator MSP the polar cap 
area at the crust surface becomes 

\begin{equation}
A_{pc} \sim \pi \frac{\Omega R}{c} (R \Delta)^2 \sim 10^{-2}~ \rm{km}^2~~,
\end{equation}

\noindent
about two orders of magnitude less than the conventionally assumed polar cap area $\pi \Omega R^3/c \sim 1~ \rm{km}^2$ when $P \sim$ several ms.   Such very small $A_{pc}$ are consistent with those reported by Zavlin (\cite{Zavlin06}) for the almost aligned rotator J0437, and two orthogonal rotators, J0030 and J2124.   There are, however, still
considerable uncertainties in inferring observed $A_{pc}$ from X-ray observations.     For an orthogonal dipole on the spin-axis at the crust core interface the surface polar cap area is much larger than the $A_{pc}$ of Eq. (6).   However, MSPs are identified by observations of their radio emission beams, a nearly aligned one from PSR 0437, and an expected gravitationally elongated (in latitude) fan beam from a rotating aligned dipole.   An aligned MSP is, therefore, usually observed from a direction nearly normal to its polar cap.   The observing angle would generally be much closer to tangential  in the orthogonal case.    Then the polar cap area inferred from its thermal X-ray emission would be near that of Eq (6) for PSR 0437 and a smaller projection of a larger  $A_{pc}$
of an orthogonal rotator whose dipole is on the spin-axis very near the stellar surface.    Despite this ambiguity,  support for a predicted very large reduction in observed polar cap areas for strongly spun-up MSPs seems quite strong.

\section{Pulsar spin-period glitches from spin-induced $B$-field changes}

Moving core flux-tubes continually build up shearing stress in the conducting crust which anchors $B$-field that traverses it.   If this stress grows to exceed the crust's yield strength, subsequent relaxation may, at least partly,  be through relatively sudden crustal readjustments (``crust-breaking").  Such events would cause very small 
spin-up jumps in spinning-down NSs (spin -period ``glitches").   The Sect. 2--3 model for the evolution of a core's flux-tube array suggests glitch details in pulsars similar to those of the two observed glitch families:   Crab-like glitches (C) and the very much larger giant Vela-like ones (V) of Fig. 17.

\begin{figure}
\includegraphics[width=8.8cm]{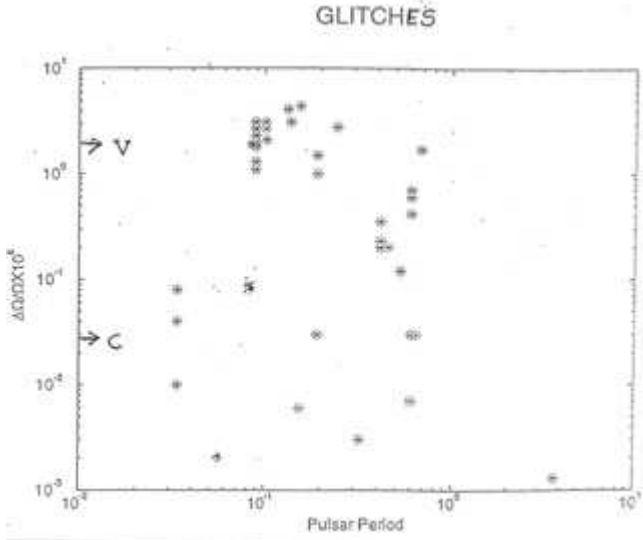}
\caption{Observed jumps (``glitches") in pulsar spin-rates ($\Delta \Omega/\Omega$) of pulsars with various periods ($P$).  The vela-like family (V) has  $\Delta \Omega/\Omega \sim 10^{-6}$.   The Crab-like one (C) has  ($\Delta \Omega/\Omega) \sim 10^{-7}  - 10^{-8}$ (19; 16; 8; 34; 7).}                                     
\end{figure}

\begin{enumerate}
\item[a)]  {\em Crab-like glitches} -- In both the $(A \rightarrow C)$ and $(C \rightarrow D)$ segments of Fig. 14, an expanding quasi-uniform vortex-array carries a flux-tube array outward with it.   If growing flux-tube-induced stress on the crust is partly relaxed by ``sudden" outward crust movements (of magnitude s) where the stress is strongest (with density preserving backflow elsewhere in the stratified crust) the following consequences are expected:
\vspace*{8pt}

\item[(1)] a ``sudden" permanent increase in $\mu_\perp$, spin-down torque, and
 $\dot{\Omega}$ with $\frac{\Delta \dot{\Omega}}{\dot{\Omega}} ~ \sim~ \rm{s/R}  \sim \Delta \theta$
(strain relaxation) $^<_\sim \theta_{max} \sim 10^{-3}$.
(This is the largest non-transient fractional change in any of the pulsar observables expected from ``breaking" the crust.)   A permanent glitch-associated jump in NS spin-down rate of this sign and magnitude ($\sim 3 \times 10^{-4})$ is indeed observed in the larger Crab glitches (Fig. 18) ({Lyne et al. 1996;  Lyne et al. 1992;  Gullahorn et al. 1997;  Wong et al. 2001}).

\begin{figure}
\includegraphics[width=8.8cm]{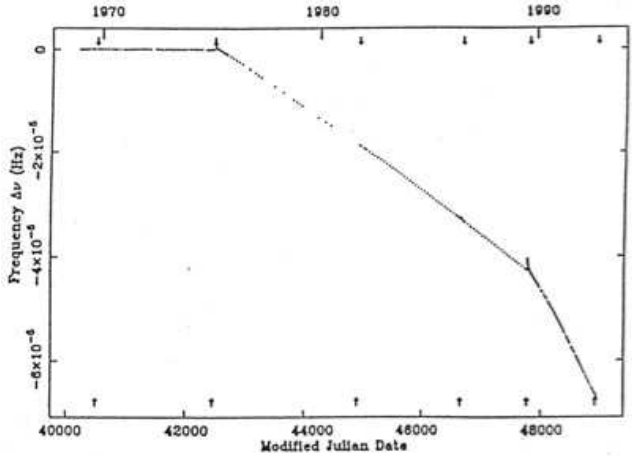}
\caption{The difference between Crab pulsar periods observed over a 23 yr interval and those predicted from extrapolation from measurement of $P$, $\dot{P}$, and $\ddot{P}$ at the beginning of that interval.   These ``sudden" permanent fractional jumps in spin-down rate  ($\Delta \dot{\Omega}/\dot{\Omega}) \sim +5 \times 10^{-4}$)  occur at glitches  ($\Delta \dot{\Omega}/\dot{\Omega} \sim 10^{-8} - 10^{-7}$)  but are $10^4$ times greater in magnitude (18; 17).}
\end{figure}

\item[(2)] a ``sudden" reduction in shear stress on the crust by the flux-tubes attached to it from below.   This is matched by an equivalent reduction in pull-back on the core's expanding vortex array by the core flux-tube array attached to it.     The n-vortices therefore ``suddenly" move out to a new equilibrium position where the Magnus force on them is reduced by just this amount.    The high density SF-n sea therefore spins down a bit.     All the (less dense) charged components of the NS (crust, core-p and-e) together with the flux-attached n-vortex-array spin-up much more.      (The total angular momentum of the NS does not change significantly in the brief time for development of the glitch.)    A new equilibrium is established in which the charged components (all that can be observed, of course, is $P$ of the crust's surface) have been spun up.   For Crab $B$ and $P$, the estimated (26) $\Delta \Omega/\Omega \sim 10^{-4} (\Delta/ \dot{\Omega})$, consistent with both the Crab glitches of Fig. 18 and also with much smaller Crab glitches not shown there ({Wong et a. 2001}).
\vspace*{8pt}

\item[b)]  {\em Giant Vela-like (V) glitches}.    The second (V)-family of glitches differs from that of Crab-like ones (C) in several ways.
\vspace*{8pt}

\noindent
\item[(1)] $\left(\frac{\Delta \Omega}{\Omega}\right)_V ~ \sim 10^2  \times \left(\frac{\Delta \Omega}{\Omega} \right)_C $
\vspace*{8pt}

\noindent
\item[(2)] V-glitches develop their $\Delta \Omega$ in less than $10^2$ sec.; the $\Delta \Omega$ of a V-glitch is already decreasing in magnitude when first resolved (16), while C-glitches are still rising toward their full $\Delta \Omega$ for almost $10^5$ sec ({McCulloch et al. 1990;  Lyne et al. 2000}).
\vspace*{8pt}

\noindent
\item[(3)] V-glitches are observed in pulsars (mainly, but not always) in Fig. 12 along ($C \rightarrow D$) while C-glitches are observed all along $(A \rightarrow C \rightarrow D)$.
\vspace*{8pt}

\noindent
\item[(4)] The C-glitch proportionality between $\Delta \dot{\Omega}/\dot{\Omega}$ and $\Delta \Omega/\Omega$ would greatly overestimate  $(\Delta \dot{\Omega}/\dot{\Omega}$ for V-glitches.
\end{enumerate}
\vspace*{8pt}

The existence of a second glitch family, with V-properties, is expected from a second effect of vortex-driven flux-tube movement in a NS core.   If there were no very dense, comoving, flux-tube environment around them, outward moving core-vortices could smoothly shorten and then disappear  as they reached the core's surface at its spin-equator.  (We ignore crustal SF-n here.)   However, the strongly conducting crust there resists entry of the flux-tubes which the vortices also bring with them to the crust's base.  This causes a pile-up of pushed flux-tubes into a small equatorial annulus (Figs. 19 and 20) which delays the final vortex-line disappearance.   The vortex movement in which they vanish occurs either in vortex-line flux-tube cut-through events, or, more likely, in a sudden breaking of the crust which has been over stressed by the increasing shear-stress on it from the growing annulus.   Giant V-glitches have been proposed as such events ({Ruderman 2004a,b})  allowing a ``sudden" reduction of part of this otherwise growing annulus of excess angular momentum and 

\begin{figure}
\includegraphics[width=4.2cm]{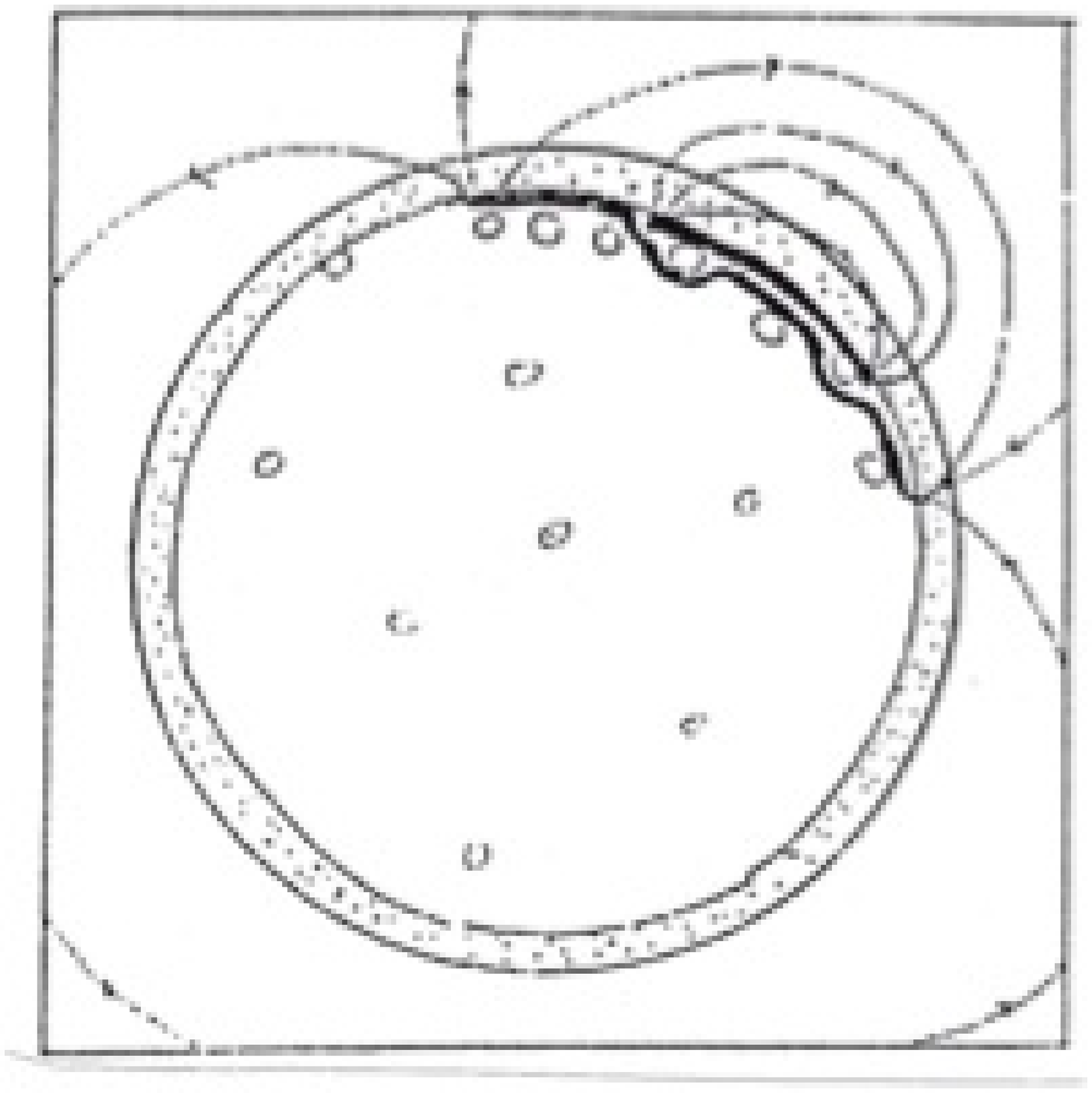}
\includegraphics[width=4.2cm]{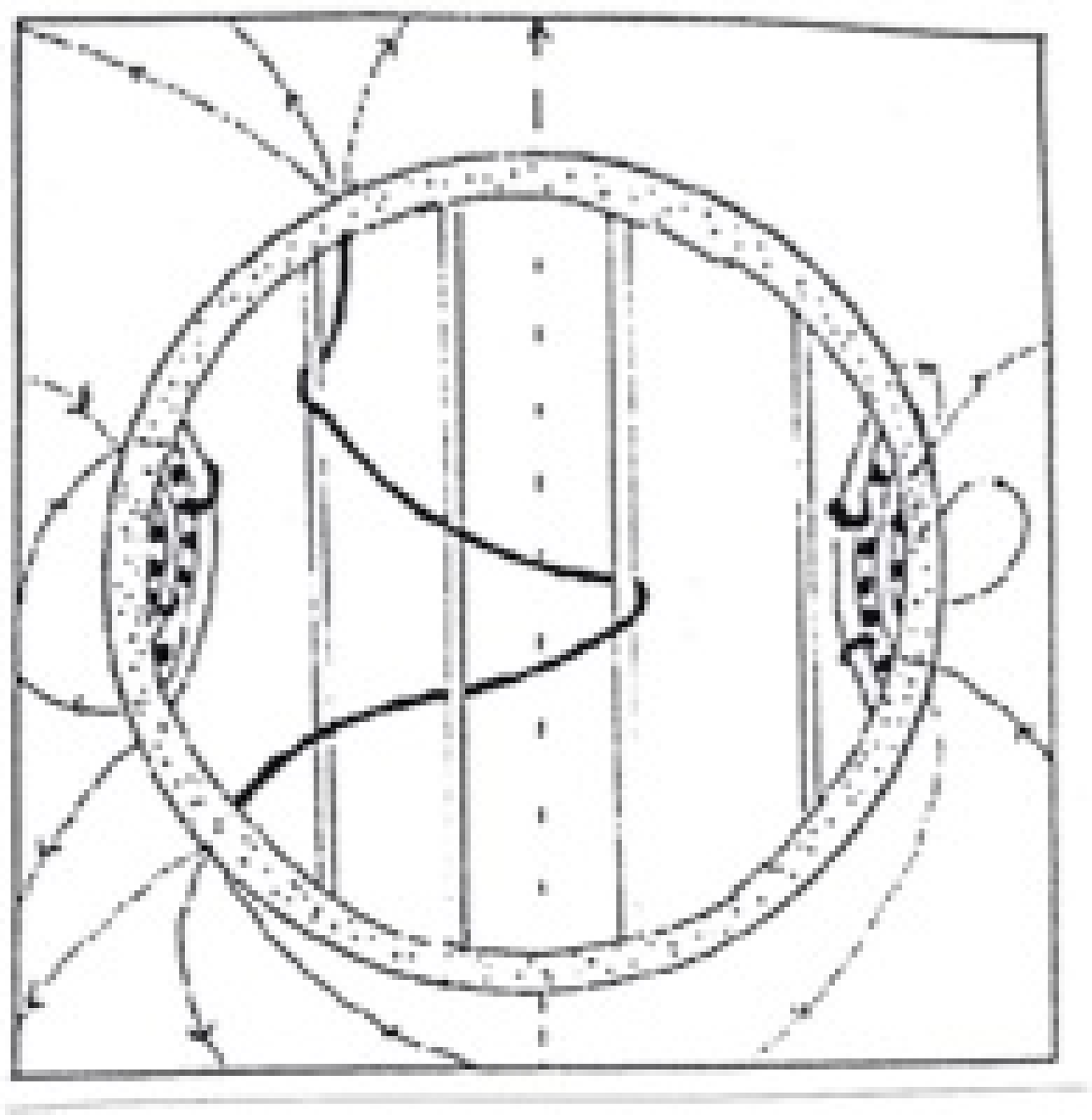}
\caption{(L) The configuration (top view) of Fig. 5b after further spin-down.   Flux-tubes are piling up in an equatorial annulus a the core-crust interface.   The blocked flux-tubes, in turn, block short segments of vortex lines which forced them into this annulus.}       
\end{figure}
\begin{figure}
\caption{(R) A side view of the representation of the Fig. 19 configuration with the addition of a flux-tube, which the expanding vortex-array has not yet forced out to a radius $\sim R$.}                                                                
 \end{figure}

\noindent
also some of the magnetic flux trapped within it.  These would not begin until enough vortex-lines, initially distributed almost uniformly throughout the core, have piled up in the annulus for the flux-tubes they bring with them to supply the needed shear stress.   Estimates of V-glitch $\Delta \dot{\Omega}/\dot{\Omega}$ magnitudes
are less reliable than those for C-glitch ones.   A very rough one, based upon plausible guesses and an assumed
$\mu/R$ about the same as those in the larger C-glitches, suggest V-glitch repetition rates and magnitudes similar to observed ones 

\section{Help wanted}

Three areas  stand out as needing much more study.

\begin{enumerate}
\item[1)] The origin of a NS's magnetic field --     Speculations include
(a) conservation during a NS's violent birth of flux already inside its ancestor.   NS fluxes are comparable to those in magnetic White Dwarfs, and  the toroidal field within the sun; (b) short-lived post-partum dynamos ({Thompson \& Duncan 1996});   (c) field amplification in asymmetric supernova explosions; (d) toroidal field breakout after wind-up from different rotation
imparted at birth ({Ruderman et al. 2000}); (e) thermoelectric generation ({Applegate et al. 83}); and (f) exterior field reduciton from burial by fall-back of some of the initially exploded matter.
\vspace*{8pt}

\item[2)] Prehistory -- The proposed spin-down biography of a NS surface $B$ presented in Sect. 2 began at $A$ (or perhaps $A^\prime$) in Fig. 14 when that typical NS is expected to be about $10^3$ yrs old.   Before that its crust had solidified (age $\sim$ a minute), its core protons had become superconducting $(\sim$ 1 yr?), and core neutrons
became superfluid ($\sim 10^3$ yrs?).   If so, there would be a nearly $10^3$ year interval between formation 
of the NS core's magnetic flux-tube array and control of that array's movement by a SF-n vortex array.   During that interval an early magneto-hydrodynamic equilibrium involving poloidal and toroidal fields, and some crustal shear stress could be upset.   Dramatically altered $B$-field stresses after flux-tube formation could induce movements in the overstrained crust which cause $B_d$ to change ({Ruderman 2004a; Ruderman 200b}).  Recent reconsideration of drag on moving flux-tubes  
({Jones 2006}) suggests the core flux-tube adjustment might take $\sim 10^3$ yrs.   For many NSs, depending on details of their initial $B$ structure, dipole moments could become much smaller in that $10^3$ yr epoch.
Post-partum $B_d$ values and subsequent drops in their sizes have been estimated and proposed  (\cite{Ruderman04a, Ruderman04b}) as the reason many young pulsars have spin-down ages $(P/2\dot{P})$ up to $10^2$ times greater than their true ages (Fig. 21).
\vspace*{8pt}

\begin{figure}
\includegraphics[width=8.8cm]{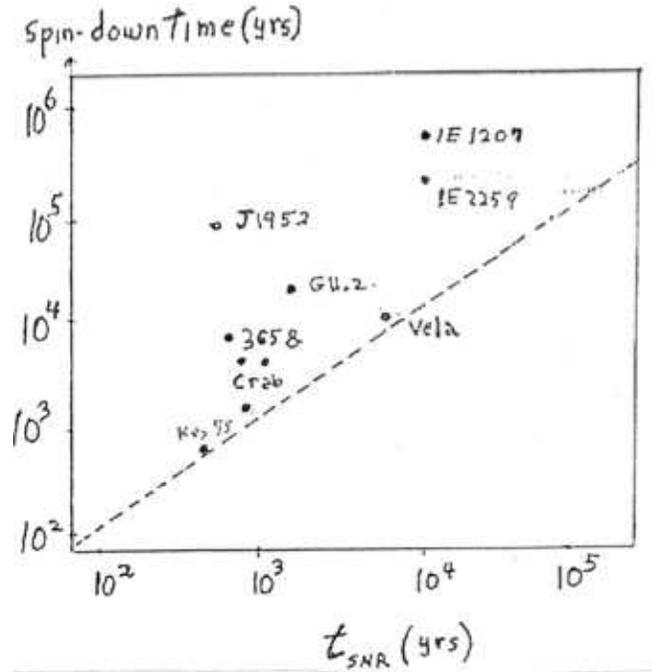}
\caption{Observed spin-down times for pulsars ($P/2\dot{P}$) vs the time since birth of these same pulsars as inferred by the ages of the supernova remnants in which they are still embedded ($t_{SNR}$).}                                                                            
\end{figure}

If present $P < 10^{-1} \rm{s}$  it may not be implausible that such pulsars simply were born with a spin-period near that observed now.   In others (e.g. 1E 1207 with $P \sim$ 0.4 s) the case for an earlier epoch with very much larger $B_d$ may be more attractive.  It has been suggested by Geppert and others that this might have been caused by burial of surface field by slow fall-back.
\vspace*{8pt}

A second consequence of a very early, large reduction in $B_d$ of some NSs would be a comparable reduction in their $A_{pc}$ to far below the canonical $\pi \Omega R^3/c$ (cf Eq (5)).
The $A_{pc}$  of Sect. 5 would then be small because presently observed pulsars ``remember" their early very large $B_{pc}$ but not the $B_d$  they then had.   Subsequent changes in $B_d$ and Eqs (1) and (2) would need reconsideration.
\vspace*{8pt}

An early epoch between the formation of SCp flux-tubes and SFn vortices arrays is rich in possibilities and needs further exploration.
\vspace*{8pt}

\item[3)] Precession -- Large, long lasting and long period ($\sim$ yr) free precession of a NS appears to be incompatible with simple, canonical neutron star models ({Shaham 1977;  Sedrakian et al. 1999; Link 2003}).
This is especially the case for the model considered above in which the NS core's n-SF vortex lines are ``tied" to the NS crust by magnetic flux which interacts strongly with both.
However, there are a substantial number of observations of significant long period ($\sim$ year) oscillations around expected radio-pulse arrival times (eg ({Lyne et al. 2000})  which have been interpreted as evidence for large amplitude free precession of NSs).    How to modify the canonical model of a  spinning-down (-up) NS to allow such sustained free precession and also preserve the consequences described above for magnetic fields and glitches seems a severe and crucial problem (cf. ({Link 2003; Alpar 2005}).   One proposed resolution ({Ruderman \& Gil 2006})  is that the observations attributed to free precession of the NS are from  slow oscillations in emission beam direction and spin-down torque coming from a  very slow ``drifting" (precession) only of the current-pattern within  a polar cap accelerator.
\end{enumerate}

\section{Acknowledgments}

I am happy to thank A. Beloborodov, E.V. Gotthelf, J.P. Halpern, P. Jones, A. Lyne, J. Sauls, J. Trumper, and colleagues at the Institute of Astronomy (Cambridge) and the Center for Astrophysics and Space Sciences (UCSD) for helpful discussions and hospitality.
\vspace*{8pt}

Much of this paper is an expanded version of one presented at the August 2005 meeting in Amsterdam to celebrate E. van den Heuvel.   Those Proceedings will be published by Elsevier Science (R. Wijers et al. eds.).


\end{document}